\documentclass[12pt]{iopart}

\usepackage{graphicx}% Include figure files
\usepackage{dcolumn}% Align table columns on decimal point
\usepackage{bm}% bold math
\usepackage{verbatim}

\begin{document}

\title[Optimal interlayer hopping and high $T_C$ BEC in quasi 2D superconductors]{Optimal interlayer hopping and high temperature Bose-Einstein condensation of local pairs in quasi 2D superconductors}
       
\author{P E Kornilovitch}
\address{Hewlett-Packard Company, Corvallis, OR 97330, USA} 
\address{Department of Physics, Oregon State University, Corvallis, OR 97331, USA} 
\ead{pavel.kornilovich@hp.com}

\author{J P Hague}
\address{Department of Physical Sciences, The Open University, Walton Hall, Milton Keynes, MK7 6AA, UK} 
\ead{Jim.Hague@open.ac.uk}

\date{\today}  % It is always \today, but any date may be explicitly specified

\begin{abstract}

Both FeSe and cuprate superconductors are quasi 2D materials with high transition temperatures and local fermion pairs. Motivated by such systems, we investigate real space pairing of fermions in an anisotropic lattice model with intersite attraction, $V$, and strong local Coulomb repulsion, $U$, leading to a determination of the optimal conditions for superconductivity from Bose-Einstein condensation. Our aim is to gain insight as to why high temperature superconductors tend to be quasi 2D. We make both analytically and numerically exact solutions for two body local pairing applicable to intermediate and strong $V$. We find that the Bose Einstein condensation temperature of such local pairs is maximal when hopping between layers is intermediate relative to in-plane hopping, indicating that the quasi 2D nature of unconventional superconductors has an important contribution to their high transition temperatures.

\end{abstract}

     % PACS, the Physics and Astronomy Classification Scheme
%\submitto{\JPCM}

\maketitle

\section{\label{UV3d:sec:one}
Introduction
}

The recent discovery of layered FeSe based superconductors \cite{Dagotto2013,He2013} has reinforced the view that the quasi 2D structure of high temperature superconductors is important to obtain significant transition temperatures. However, this leads to a conundrum: a small interlayer hopping, $t_{\perp}$, destroys 3D coherence of pair motion such that a purely two dimensional system should not superconduct. Conversely, a large $t_{\perp}$ increases the kinetic energy to the point that pairs are destroyed and $T_c$ goes down as a result. The focus of this work is the effect of interlayer hopping $t_{\perp}$ on the condensation temperature, $T_c$, of local fermionic pairs formed in a model of intersite attraction, $V$, and local Coulomb repulsion, $U$. The question that we seek to answer with the present paper is whether there is an optimal anisotropy associated with particle transport that leads to an enhancement in the transition temperature at moderate interlayer hopping. The existence of such a maximum in the transition temperature would point to anisotropy as a key feature in high $T_c$ supreconductors. In the following, it is shown that $T_c$ is maximal when the underlying single fermion spectrum is strongly anisotropic.

A feature of FeSe and cuprate superconductors is believed to be the formation of local pairs above the superconducting transition temperature, that then Bose condense to form a superconductor \cite{Dagotto2013}. Since these pairs are formed above the transition temperature they are often known as preformed pairs. The existence of local pairs above the transition temperature provides an intuitive explaination of a number of normal and superconducting properties of the cuprates, including the pseudogap~\cite{Alexandrov1993,Alexandrov2011}. Preformed pair superconductors form in the following way~\cite{Ogg1946,Schafroth1954,Schafroth1957,Bogoliubov1970,Micnas1990,Alexandrov1994}: (i) The attractive interaction between two charged fermions is sufficiently strong to form a bound state in the absence of a Fermi surface. (ii) At low carrier densities, the fermion pairs behave approximately as bosons and undergo Bose-Einstein condensation (BEC) as temperature is dropped below the transition temperature. Macroscopically, BEC of pairs manifests itself as superconductivity. (iii) At high densities, the pairs begin to overlap and the fermions experience additional repulsion due to the exclusion principle. Such `overcrowding'~\cite{Salje1984,Kyung1998} gradually evaporates pairs as the density increases. Superconductivity either transitions to the BCS regime or is destroyed altogether. 

In a quasi 2D system, the condensation temperature of an anisotropic ideal Bose gas in {\em continuous} space is given by
\begin{equation}
k_B T_{c, \: {\rm cs}} = 3.31 \hbar^2 \frac{\nu^{2/3}_{b}}{(m_{bx} m_{by} m_{bz})^{1/3}} \: , 
\label{3UV2d:eq:one}
\end{equation}
which also applies to local pairs of fermions if the pairs are well separated. Here $\nu_b$ is the volumetric density of bosons and $m_{bi}$ is the $i$-th component of the boson's effective mass. Since $T_c$ is an increasing function of $\nu_b$, it can be systematically raised by increasing $\nu_b$ until the bosons are close packed. Thus $T_c$ is maximal when $\nu_b$ is approximately equal to an inverse volume of the boson $\Omega_b$. Such a maximal $T_c$ will be referred to as the {\em close packed} condensation temperature and denoted $T^{\ast}_c$. In continuous space, it is given by $k_B T^{\ast}_{c, \: {\rm cs}} = 
3.31 \hbar^2 A_{\rm cs}/\Omega^{2/3}_b (m_{bx} m_{by} m_{bz})^{1/3}$,
where $A_{\rm cs}$ is a dimensionless constant $\sim 1$ that accounts for the approximate nature of the argument. This formula can also be applied to lattice fermion pairs that behave approximately as hard core bosons. The coefficient is different but the dependence on the pair volume $\Omega_p$ and pair effective masses $m_{px} = m_{py} \equiv m$ and $m_{pz} \equiv m_{\perp}$, is the same
\begin{equation}
k_B T^{\ast}_{c, \: {\rm lat}} = \hbar^2 
\frac{A_{\rm lat}}{\Omega^{2/3}_p (m^2 m_{\perp})^{1/3}} \: . 
\label{3UV2d:eq:three}
\end{equation}

\begin{figure}[t]
\begin{center}
\includegraphics[width=0.60\textwidth]{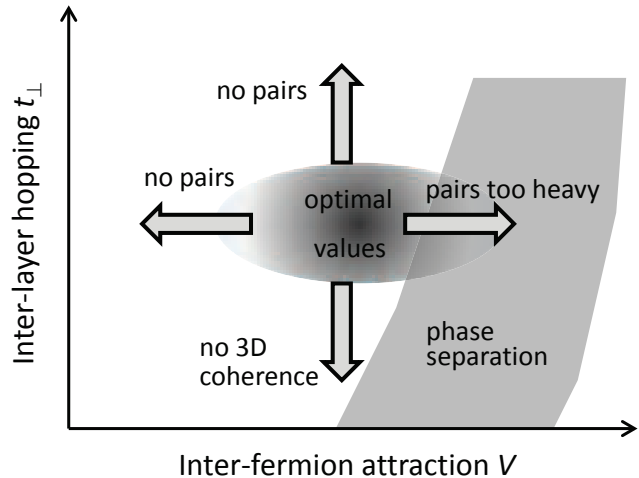}
\caption{A generic phase diagram of fermion models with core repulsion and finite-radius attraction. If $V$ is too small then pairing does not occur. For large $V$ pairs become heavy, and eventually phase separate destroying superconductivity. If $t_{\perp}$ is too small, then 3D coherence and therefore superconductivity is lost. Large $t_{\perp}$ leads to excess kinetic energy that also breaks pairs. Therefore an optimal superconductor might be expected to have intermediate $V$ and $t_{\perp}$.}
\label{UV3d:fig:one}
\end{center}
\end{figure}

Lattice bound states have two key properties. (i) At large binding energies $\Delta$, the effective pair mass scales~\cite{Mattis1986} as $m, m_{\perp} \propto \Delta$. (ii) The volume of a fermion pair on a lattice, $\Omega_p$, must approach one lattice cell at strong coupling $\Delta \rightarrow \infty$. Conversely, the volume diverges at weak coupling: $\Omega_p \rightarrow \infty$ as $\Delta \rightarrow 0$.  Realistic models for local pairing comprise a short-range repulsive part originating from direct Coulomb repulsion and a mid-range weak attractive part~\cite{Schafroth1957} which is thought to result from overscreening of the Coulomb tails by phonons, magnons or other mediating excitations. The simplest potential of such kind that still allows (partial) analytical treatment consists of a strong on-site Hubbard repulsion $U$ and an {\em instantaneous} nearest neighbor attraction $V$. Hereafter, such a potential will be called the $UV$ model. Consider the effect of $V$ on $T^{\ast}_{c}$. (From now on, the subscript `lat' will be omitted.) At small $V$ close to the binding threshold, the effective pair volume $\Omega_p$ diverges, and $T^{\ast}_{c} \rightarrow 0$. In the opposite limit of large $V$, the pair volume tends to a constant whereas all the masses scale as $\propto V$. As a result, $T^{\ast}_{c} \propto 1/V \rightarrow 0$. Thus $T^{\ast}_{c}$ tends to zero at both small and large $V$, which implies a maximum at some intermediate optimal $V$. Likewise, there is an optimal interlayer hopping $t_{\perp}$. A large $t_{\perp} \approx t$ increases the overall kinetic energy and may destroy the pairs (at small or intermediate $V$). Thus a large $t_{\perp}$ is analogous to a weak attraction, and $T^{\ast}_{c} \rightarrow 0$. (The quasi-1D case $t_{\perp} > t$ is not a subject of this paper.) Conversely, a very small $t_{\perp}$ produces a large interlayer mass $m_{\perp}$ and destroys 3D coherence. Again, a vanishing condensation temperature results. Therefore, one might expect to arrive at a generic phase diagram shown in figure~\ref{UV3d:fig:one}. The purpose of the current work is to demonstrate the features of this phase diagram in a microscopic model.

Let us address applicability of the ideal BEC formula (\ref{3UV2d:eq:three}) to dense systems. First, we note that even a strong boson-boson repulsion does not affect the critical temperature very much. Even in superfluid $^4$He, which is a strongly interacting system, the critical temperature is only about 30\% less than the BEC value. Within our approach, such an uncertainty is easily absorbed into the uncertainty of $A_{\rm lat}$. Secondly, $A_{\rm lat}$ will be set to be smaller than 1, for reasons outlined in section~\ref{UV3d:sec:threefour}. Selecting a small value for $A_{\rm lat}$ is effectively the same as diluting the system, thereby reducing the pair-pair interaction. Since the wave function decays exponentially, we estimate that the inter-pair interaction is of the order 1\% or less of the total intra-pair interaction strength for all values of $A_{\rm lat}$ used in this paper. Exact derivation of the inter-pair potential requires solving a four-fermion problem, which goes beyond the capabilities of the present method. At the same time, the {\em sign} of the effective interaction is important for the phenomenon of phase separation discussed here. It is known from analysis of three fermions that a fermion pair repels a third fermion in a pure 2D system~\cite{Kornilovitch2014}. The repulsion is due to the exclusion principle acting between a pair member and the extra fermion. This argument, being qualitative, should remain valid for four fermions (two boson pairs) in a quasi-2D system. One can speculate that the pair-pair interaction is repulsive at least in a finite parameter region above the pairing threshold.        

The work presented in this paper goes beyond previous work by presenting exact analytical and numerical solutions for local pairing in quasi-2D systems, and is to our knowledge the first detailed study using exact results to study the effect of anisotropy on local pairing and superconductivity. The paper is organised as follows: In section~\ref{UV3d:sec:three} we analyze the problem in detail within a phenomenological lattice model that includes anisotropic hopping $t_{\perp}$, on-site Hubbard repulsion $U$ and intersite attraction $V$: the tetragonal $UV$ model. Exact expressions for two-fermion energy, dispersion, wave function, and effective radius are combined to derive the conditions of pair close packing. Numerical results including the model phase diagram are given in section~\ref{UV3d:sec:four}, where we also present a Quantum Monte Carlo analysis of two fermions coupled by a retarded interaction mediated by phonons, showing that models with retardation effects also have a peak in $T_c$ as $t_{\perp}$ varies. Finally, a summary and conclusions are given in section~\ref{UV3d:sec:six}.

\section{\label{UV3d:sec:three}
Tetragonal $UV$ Model
}

In this section, we make exact calculations for the properties of local pairs in an {\em anisotropic} $UV$ model. The $UV$ model can be derived from a number of interactions, such as the electron-phonon interaction, electron-spin wave interaction and $t-J$ model, as has been extensively discussed in the literature~\cite{Micnas1990,Scalapino2012}. In the model studied here, hopping and interaction within an $xy$-plane is assumed isotropic. The difference with previous work is that both bare fermion hopping and nearest-neighbor attraction between the layers may differ from corresponding in-plane values. To reflect lattice symmetries, the model will be referred to as the {\em tetragonal} $UV$ model. The main calculation tool in this section is the equality of a Bose integral to an inverse pair volume
\begin{equation}
\frac{A}{\Omega_p} = 
\int_{\rm BZ} \frac{d^3 {\bf K}}{(2\pi)^3} 
\frac{1}{\exp{\left\{ \frac{E({\bf K}) - E_0}{k_B T^{\ast}_c} \right\} } - 1} \: , 
\label{3UV2d:eq:four}
\end{equation}
where again $A$ is a constant that is less than 1. [Note that the constant $A$ is slightly different from $A_{\rm lat}$ since it appears directly above $\Omega$, rather than $\Omega^{2/3}$.] $E_0$ is the bottom of the pair energy band, that is the value that the pair chemical potential assumes at condensation (a similar integral is used to compute the expressions in (\ref{3UV2d:eq:one}) and (\ref{3UV2d:eq:three})). Momentum components $K_i$ are measured in inverse lattice constants and pair volume $\Omega_p$ is measured in unit cell volumes. Equation (\ref{3UV2d:eq:four}) requires knowledge of the entire pair dispersion $E({\bf K})$, pair wave function $\psi$, as well as calculation of the effective pair radius and Bose integral. All operations are nontrivial and described in detail in the following subsections.   
  
\begin{comment}
We note that the constant, $A$ is slightly different from $A_{\rm
  lat}$: since it appears directly above $\Omega$, rather than
$\Omega^{2/3}$, the scaling for the inter-particle interaction,
$r_{\rm inter}\sim \langle R \rangle / A^{1/3}$. As such, the size of inter-pair
interactions is estimated to be less than approximately 1\% for $A<0.1$.
\end{comment}

\subsection{\label{UV3d:sec:threeone}
Hamiltonian
}

The Hamiltonian of the tetragonal $UV$ model is given by    
\begin{eqnarray}
H  & = & - t \sum_{{\bf m}, {\bf b}, \sigma} 
             c^{\dagger}_{{\bf m} \sigma} c_{{\bf m} + {\bf b}, \sigma}   
         - t_{\perp} \sum_{{\bf m}, {\bf b}_{\perp}, \sigma} 
             c^{\dagger}_{{\bf m} \sigma} c_{{\bf m} + {\bf b}_{\perp}, \sigma}   
\nonumber \\    
    &  & + \frac{U}{2} \sum_{\bf m} \hat{n}_{\bf m} \left( \hat{n}_{\bf m} - 1 \right) + H_{V} \: ,     
\label{3UV2d:eq:five} \\    
H_{V} & = & - \frac{V}{2} \sum_{{\bf m}, {\bf b}} \hat{n}_{\bf m} \hat{n}_{{\bf m} + {\bf b}} 
            - \frac{V_{\perp}}{2} \sum_{{\bf m}, {\bf b}_{\perp}} 
                  \hat{n}_{\bf m} \hat{n}_{{\bf m} + {\bf b}_{\perp}} \: . 
\label{3UV2d:eq:fiveone}
\end{eqnarray}
Here, $c^{\dagger}$ and $c$ are spin-$\frac{1}{2}$ fermion creation and annihilation operators, {\bf m} numbers lattice sites, ${\bf b} = \left( \pm {\bf x}, \pm {\bf y} \right)$ numbers the four nearest neighbors within the $xy$ plane, ${\bf b}_{\perp} = \pm {\bf z}$ are the two nearest lattice neighbors across the planes, $\sigma = \pm \frac{1}{2}$ is the $z$-axis spin projection, and $\hat{n}_{\bf m} = \sum_{\sigma} c^{\dagger}_{{\bf m} \sigma} c_{{\bf m} \sigma}$ is the total fermion number operator on site ${\bf m}$. The kinetic energy is defined by in-plane and between-the-planes hopping amplitudes $t$ and $t_{\perp}$. The ratio $t_{\perp}/t$ defines anisotropy of the one-particle dispersion
\begin{equation}
\varepsilon({\bf k}) = 
- 2t \left( \cos{k_x} + \cos{k_y} \right) - 2 t_{\perp} \cos{k_z} \: ,  
\label{3UV2d:eq:six}
\end{equation}
where ${\bf k}$ is the one particle momentum, and $k_{x}$, $k_{y}$ and $k_{z}$ are its components. Inter-fermion interaction is defined by three parameters: on-site (Hubbard) repulsion $U$, in-plane nearest-neighbor attraction $V$, and inter-plane nearest-neighbor attraction $V_{\perp}$. (Note that the Hubbard term is equivalent to its other standard form $U n_{{\bf m} \uparrow} n_{{\bf m} \downarrow}$.) More complex and realistic forms of attractive potential $H_V$ -- longer range and of variable strength -- are extremely difficult to treat by exact methods~\cite{Kornilovitch1995}. Thus the $UV$ interaction of (\ref{3UV2d:eq:five}) may be regarded as the simplest pseudopotential that leads to the phase diagram of figure~\ref{UV3d:fig:one}. 

If $t_{\perp} = 0$ and $V_{\perp} = 0$, (\ref{3UV2d:eq:five})-(\ref{3UV2d:eq:fiveone}) reduce to the 2D square $UV$ model that has been studied by several authors~\cite{Lin1991,Petukhov1992,Kornilovitch2004}. For the purposes of this paper it is useful to recall the $s$-pair binding condition: at zero total momentum ${\bf K} = (0,0)$, two fermions bind into an $s$-wave bound state when
\begin{equation}
V > V_{s, \, sq} = \frac{2 U t}{U + 8t} \: .  
\label{3UV2d:eq:seven}
\end{equation}
In the opposite limit, $t_{\perp} = t$ and $V_{\perp} = V$, (\ref{3UV2d:eq:five}) reduces to the isotropic 3D $UV$ model on the simple cubic lattice~\cite{Micnas1990,Davenport2012}. For zero pair momentum, the $s$-wave binding threshold can be expressed via a Watson integral. Expanding the previous results (see section 3.2 in \cite{Micnas1990} and appendix in \cite{Davenport2012}), one obtains  
\begin{equation}
V > V_{s, \, sc} = \frac{ 24 t^2 ( 1 + U M_{000} ) }{( U + 12t )( 12t M_{000} - 1 ) } \: ,  
\label{3UV2d:eq:sevenone}
\end{equation}
where $M_{000}$ is defined below in (\ref{3UV2d:eq:fourteen}). For $a = b = c = 4t$ and $\vert E \vert = 12 \, t$, $M_{000} = ( 7.91355 \, t )^{-1}$.  

The tetragonal $UV$ model, (\ref{3UV2d:eq:five})-(\ref{3UV2d:eq:fiveone}), has not been considered before. In the next subsection, a general solution for two-fermion states is derived. Most of the numerical results presented later in section~\ref{UV3d:sec:four} will be limited to $V_{\perp} = 0$. This is justified on physical grounds, since most low dimensional superconductors have much larger lattice constants in the direction perpendicular to the planes. For this reason, most preformed pair mechanisms of superconductivity assume pairing within planes and no pairing between planes.

\subsection{\label{UV3d:sec:threetwo}
Two-fermion energies and wave functions
}

The general procedure of solving two-fermion problems in isotropic $UV$ models has been described elsewhere, see, e.g., \cite{Kornilovitch1995,Kornilovitch2004,Davenport2012}. The present treatment has two novel elements. First, singlet and triplet states are separated from the start. That reduces the eigenvalue matrix from $7 \times 7$ to $4 \times 4$, which simplifies numerical calculations. Second, the wave function is computed explicitly, as it is needed for evaluation of the pair's effective volume. For those reasons, the method is outlined below.

Let $\psi_{+}({\bf k}_1,{\bf k}_2)$ be a two-particle wave function in momentum space, symmetrized with respect to permutation of ${\bf k}_1$ and ${\bf k}_2$. The two-particle Schr\"odinger equation can be written as 
\begin{eqnarray}
\psi_{+}({\bf k}_1,{\bf k}_2) = & & \frac{1}{ E - \varepsilon({\bf k}_1) - \varepsilon({\bf k}_2) } 
\times 
\nonumber \\
& & \times \left\{ U \Phi_{\bf 0}({\bf K})  
                 - V \Phi_{\bf x}({\bf K})
                 [ e^{-i {\bf k}_1 {\bf x}} + e^{-i {\bf k}_2 {\bf x}} ] \right.
\nonumber \\
& & -        V         \Phi_{\bf y}({\bf K}) [ e^{-i {\bf k}_1 {\bf y}} + e^{-i {\bf k}_2 {\bf y}} ]
\nonumber \\
& & - \left. V_{\perp} \Phi_{\bf z}({\bf K}) [ e^{-i {\bf k}_1 {\bf z}} + e^{-i {\bf k}_2 {\bf z}} ] 
\right\} ,
\label{3UV2d:eq:eight}
\end{eqnarray}
where
\begin{equation}
\Phi_{\bf 0}({\bf K}) = \frac{1}{N} \sum_{\bf q} \psi_{+}({\bf q},{\bf K}-{\bf q}) \: ,  
\label{3UV2d:eq:nine}
\end{equation}
\begin{equation}
\Phi_{\bf x}({\bf K}) = \frac{1}{N} \sum_{\bf q} \psi_{+}({\bf q},{\bf K}-{\bf q}) \,
e^{i {\bf q} {\bf x}} \: ,  
\label{3UV2d:eq:ten}
\end{equation}
\begin{equation}
\Phi_{\bf y}({\bf K}) = \frac{1}{N} \sum_{\bf q} \psi_{+}({\bf q},{\bf K}-{\bf q}) \,
e^{i {\bf q} {\bf y}} \:  ,  
\label{3UV2d:eq:eleven}
\end{equation}
\begin{equation}
\Phi_{\bf z}({\bf K}) = \frac{1}{N} \sum_{\bf q} \psi_{+}({\bf q},{\bf K}-{\bf q}) \,
e^{i {\bf q} {\bf z}} \: ,  
\label{3UV2d:eq:twelve}
\end{equation}
$N$ is the number of lattice sites, and ${\bf K} = {\bf k}_1 + {\bf k}_2$ is the total lattice momentum of the two fermions. Notice that after symmetrization it is sufficient to introduce only four integral functions $\Phi$, rather than seven when wave functions are not symmetrized~\cite{Davenport2012}. Substitution of (\ref{3UV2d:eq:eight}) into (\ref{3UV2d:eq:nine})-(\ref{3UV2d:eq:twelve}) results in a set of four algebraic equations on $\Phi_{i}({\bf K})$ which, after transformations, can be cast in an eigenvalue form
\begin{equation}
%\begin{center}
\hspace{-4.0cm}
\scriptstyle{
\left( \begin{array}{c}
             \Phi_{\bf 0}  \\
e^{-i K_x/2} \Phi_{\bf x}  \\
e^{-i K_y/2} \Phi_{\bf y}  \\
e^{-i K_z/2} \Phi_{\bf z}
\end{array} \right) 
= \left( \begin{array}{cccc}
-U M_{000} & 2 V M_{100}               & 2 V M_{010}               & 2 V_{\perp} M_{001} \\
-U M_{100} &   V ( M_{000} + M_{200} ) & 2 V M_{110}               & 2 V_{\perp} M_{101} \\
-U M_{010} & 2 V M_{110}               &   V ( M_{000} + M_{020} ) & 2 V_{\perp} M_{011} \\
-U M_{001} & 2 V M_{101}               & 2 V M_{011}               &   V_{\perp} ( M_{000} + M_{002} )  
\end{array} \right) 
\cdot
\left( \begin{array}{c}
             \Phi_{\bf 0}  \\
e^{-i K_x/2} \Phi_{\bf x}  \\
e^{-i K_y/2} \Phi_{\bf y}  \\
e^{-i K_z/2} \Phi_{\bf z}
\end{array} \right) .
}
\label{3UV2d:eq:thirteen}
%\end{center}
\end{equation}
The matrix elements $M_{nml}$ are given by generalized Watson integrals~\cite{Zucker2011}
\begin{equation}
M_{nml} \equiv \frac{1}{N} \sum_{\bf q} \frac{\cos{nq_x} \cos{mq_y} \cos{lq_z}}
{\vert E \vert - a \cos{q_x} - b \cos{q_y} - c \cos{q_z}} \: ,
\label{3UV2d:eq:fourteen}
\end{equation}
where $a \equiv 4 t \cos{(K_x/2)}$, $b \equiv 4 t \cos{(K_y/2)}$, and $c \equiv 4 t_{\perp} \cos{(K_z/2)}$. 

Equations~(\ref{3UV2d:eq:thirteen}) and (\ref{3UV2d:eq:fourteen}) have wider applicability than the tetragonal $UV$ model. If the potentials $V$ appearing in the second and third columns of (\ref{3UV2d:eq:thirteen}) are not equal, and the hopping amplitudes $t$ appearing in $a$ and $b$ are not equal, then the equations provide a solution to the {\em orthorhombic} $UV$ model in which both hopping and attraction strength have different values along all three coordinate axes. Likewise, (\ref{3UV2d:eq:thirteen}) remains valid for negative $U$ or positive $V$ or $V_{\perp}$ as long as the remaining attraction is strong enough to produce a bound state. (The present paper does not deal with two-fermion scattering states.) As mentioned earlier, of main physical interest here is $V_{\perp} = 0$. In this case, the $4 \times 4$ matrix in (\ref{3UV2d:eq:thirteen}) is reduced to its $3 \times 3$ top left corner. The latter acts on an eigenvector that involves only $\Phi_{\bf 0}$, $\Phi_{\bf x}$, and $\Phi_{\bf y}$, while $\Phi_{\bf z}$ remains undefined.        

Triplet $p$-wave bound states can be derived by the same procedure but in this case the starting point must be antisymmetrized basis functions $\psi_{-}({\bf k}_1,{\bf k}_2)$. Since the triplet states are not included in the thermodynamic Bose integral (for the reasons given in the next subsection) specific formulas are not presented here. Suffice it to say that the general $3 \times 3$ eigenvalue system factorizes into three independent $1 \times 1$ blocks at arbitrary ${\bf K}$. Thus the three $p$-wave pair bands do not mix throughout the entire Brillouin zone.    

The numerical sequence starts with a search for energy $E \leq - ( a + b + c )$ at which (\ref{3UV2d:eq:thirteen}) has at least one eigenvalue $\lambda = 1$. By repeating the process for all ${\bf K}$, the entire singlet dispersion is computed. This procedure requires multiple evaluation of Watson integrals $M_{nml}$, which is highly nontrivial. Direct numerical evaluation as 3D integrals is wasteful and inaccurate. A special procedure has been developed that combines two analytical integrations with one 1D numerical integration. All the necessary details are presented in \ref{UV3d:sec:appa}. Once $E$ is found for some ${\bf K}$, the pair wave function is given by (\ref{3UV2d:eq:eight}). In the last expression, $\Phi$ are eigenvector components from (\ref{3UV2d:eq:thirteen}) corresponding to the same $E$. This wave function can then be used to evaluate the pair effective radius and volume as explained in section~\ref{UV3d:sec:threefour}. 

\begin{figure}[t]
\begin{center}
\includegraphics[width=0.60\textwidth]{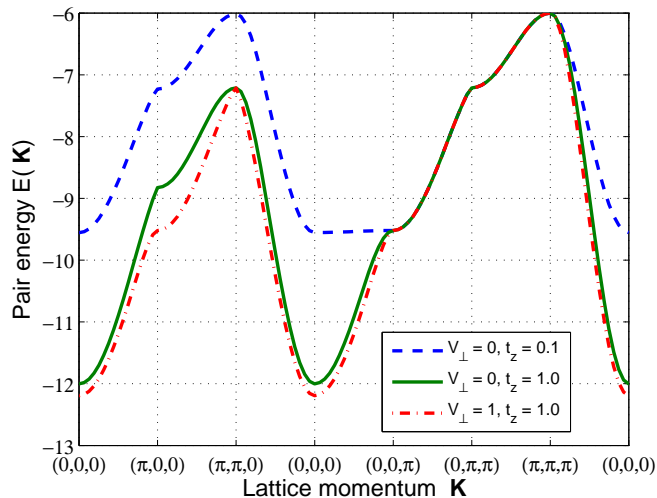}
\caption{(Color online) $s$-wave singlet pair dispersion for $U = 10$, $V = 6$. Other parameters are shown in the legend. The three dispersions are degenerate along the $(0,0,\pi)-(0,\pi,\pi)$ and $(0,\pi,\pi)-(\pi,\pi,\pi)$ directions.}
\label{UV3d:fig:two}
\end{center}
\end{figure}

An example pair dispersion is shown in figure~\ref{UV3d:fig:two}. Notice how reducing $t_z$ from 1.0 to 0.1 flattens dispersion along the $K_z$ axis. At $V_{\perp} = 0$, $t_z = 1.0$, the attraction is barely enough to produce a bound state at zero {\bf K}: the ground state energy is $E_0 = -12.00246 \, t$, computed to that level of accuracy.

\subsection{\label{UV3d:sec:threethree}
The Bose integral
}

Once a pair dispersion is known, the integral in the right hand side of (\ref{3UV2d:eq:four}) can be computed. Such integration is considered meaningful as long as a stable $s$-wave pair exists at zero ${\bf K}$. In principle, all pair branches as well as scattering two-fermion states should be included. However, those states are separated from the lowest $s$-wave branch by a finite gap, and their contributions are exponentially small at low temperatures. For very small binding energies, contribution from the scattering states becomes significant but by then the pairs are already loose and far from the optimal regimes. Thus excluding scattering states from the Bose integral does not affect the most interesting parameter region. Similarly, triplet pair branches and higher energy singlet branches ($d$-waves) are separated from the lowest branch by finite gaps. The gap is smaller at large lattice momenta but since contributions of all high-momenta states are exponentially small, excluding $p$- and $d$-wave pairs does not introduce significant numerical errors. In summary, only the lowest $s$-wave states are included in the Bose integral.   

The numerical approach to (\ref{3UV2d:eq:four}) adopted here consists of precomputing and storing $E({\bf K})$ on a fixed-step mesh within the irreducible 1/8 of the Brillouin zone and subsequent evaluation of the integral by a 3D Simpson rule. If necessary, the stored values can be used to interpolate between the mesh points by splines or any other suitable method. A major technical difficulty for the Simpson method is the singularity at zero ${\bf K}$. The singularity is integrable and as such can be isolated and treated semi-analytically. To this end, the first mesh line is shifted from zero to a small positive number $h$. As a result, the full integration volume $0 \leq K_{x,y,z} \leq \pi$ splits into 8 sub-volumes. Details of sub-volume integrations are given in \ref{UV3d:sec:appb}.  

In a typical calculation, $h = 0.01$ and the linear interval $[h,\pi]$ is divided into $q = 20$ equal steps. This requires 10,648 calculations of $E({\bf K})$ which can be achieved with reasonable computational efforts. Then $E({\bf K})$ is spline interpolated to a much denser mesh with $q$ between 120 and 200, to which the integration procedure described above is applied. The entire method has been validated by comparing with previously published results on Bose gases in simple cubic lattices: $T_c = 5.591 \, t$ for $\nu = 1$ \cite{Capogrosso2007}, $T_c = 5 \, t\nu$ \cite{Yukalov2009}, and $T_c = 5.6 \, t \nu^{0.825}$ \cite{Kleinert2014}.

\subsection{\label{UV3d:sec:threefour}
Pair effective radius and volume
}

Consider now the left hand side of (\ref{3UV2d:eq:four}), namely the effective volume $\Omega_p$ of a fermion $s$-wave pair at ${\bf K} = (0,0,0)$. Conceptually, calculation is straightforward. After solving the eigenvalue problem, (\ref{3UV2d:eq:thirteen}), for energy $E$ and eigenvector $\Phi$, the pair wave function is given by (\ref{3UV2d:eq:eight}). However, the latter equation provides $\psi$ in momentum space while $\Omega_p$ requires $\psi$ in real space. Conversion to real space involves a three dimensional integration for every value of relative coordinates $\{ \Delta {\bf r}_i \}$. Further, calculation of $\Omega_p$ is another 3D summation over real space. Thus, direct numerical evaluation of $\Omega_p$ from (\ref{3UV2d:eq:eight}) requires a 6D integration and is impractical.   

Fortunately, the specific form of the denominator in (\ref{3UV2d:eq:eight}) and single-particle dispersion, (\ref{3UV2d:eq:six}), admits analytical reduction of $\langle ( \Delta {\bf r}_i )^2 \rangle$ to combinations of 1D integrals:
\begin{equation}
\langle ( \Delta {\bf r}_i )^2 \rangle = \frac{1}{3}
\frac{\sum_{{\bf b}{\bf b}'} G_{\bf b} G_{{\bf b}'} 
         \int^{\infty}_0 d\xi \,     e^{-\xi \vert E \vert} S^{i}_{{\bf b}{\bf b}'}(\xi) }
     {\sum_{{\bf b}{\bf b}'} G_{\bf b} G_{{\bf b}'} 
         \int^{\infty}_0 d\xi \, \xi e^{-\xi \vert E \vert} T^{i}_{{\bf b}{\bf b}'}(\xi) } \: . 
\label{3UV2d:eq:twenty}
\end{equation}
Here ${\bf b} = \{ {\bf 0}, \pm {\bf x}, \pm {\bf y}, \pm {\bf z} \}$, $G_{\bf 0} = U \Phi_{\bf 0}$, $G_{\pm {\bf x}} = - V \Phi_{\bf x}$, $G_{\pm {\bf y}} = - V \Phi_{\bf y}$, $G_{\pm {\bf z}} = - V_{\perp} \Phi_{\bf z}$. The functions $S^{i}_{{\bf b}{\bf b}'}(\xi)$ and $T^{i}_{{\bf b}{\bf b}'}(\xi)$ are combinations of modified Bessel functions of various orders. Their explicit forms as well as derivation of (\ref{3UV2d:eq:twenty}) are given in \ref{UV3d:sec:appc}.
 
The next question is: once $\langle ( \Delta {\bf r}_i )^2 \rangle$ are known, what is the effective pair volume $\Omega_p$? The following conditions must be met: (i) the three coordinates enter the expression symmetrically; (ii) when any one $\langle ( \Delta {\bf r}_i )^2 \rangle$ is infinite, the volume is infinite too; (iii) when all $\langle ( \Delta {\bf r}_i )^2 \rangle$ shrink to zero, $\Omega_p \rightarrow 1$. The latter condition enforces the fact that fermionic pairs are hard-core bosons and the pair volume cannot be less than one even if the effective radius is zero. There are many choices that satisfy the above conditions. In this work, the following definition is adopted
\begin{equation}
\Omega_p = \sqrt{[ 1 + \langle ( \Delta x )^2 \rangle ] [ 1 + \langle ( \Delta y )^2 \rangle ] 
                 [ 1 + \langle ( \Delta z )^2 \rangle ] } \: . 
\label{3UV2d:eq:twentyone}
\end{equation}

Finally, one should discuss the choice of constant $A$ in the left hand side of (\ref{3UV2d:eq:four}). Admittedly, the exact condition for maximum critical temperature is not known; that would require a complete many-fermion solution that is not available yet. The factor $A$ reflects this uncertainty. $A = 1$ implies superconductivity is optimal when the average distance between pairs is equal to the $1/e$ decay distance of a pair wave function. Beyond the $1/e$ distance the wave function is not negligible, and overlap between pairs is already strong. Therefore one expects that $A$ should be somewhat less than 1. Reducing $A$ results in a sharp decrease of effective inter-pair interaction at close-packing. In a quasi-2D situation ($\triangle z \ll 1$), the mean distance between pairs scales as $1/\sqrt{A}$. Going, for example, from $A = 1.0$ to $A = 0.1$ increases the mean distance by a factor 3 and decreases density overlap by at least a factor of 100. In the following, a value of $A = 0.1$ is selected as the base point, but other values are considered as well. Variations of $A$ directly translate into variations of $T^{\ast}_{c}$ but not into variations of optimal $t_{\perp}$ and $V$ at which this $T^{\ast}_{c}$ is achieved. Although the present method cannot predict the magnitude of the critical temperature better than by order of magnitude, statements about optimal $t_{\perp}$ and $V$ are quite robust.

\begin{figure}[t]
\begin{center}
\includegraphics[width=0.60\textwidth]{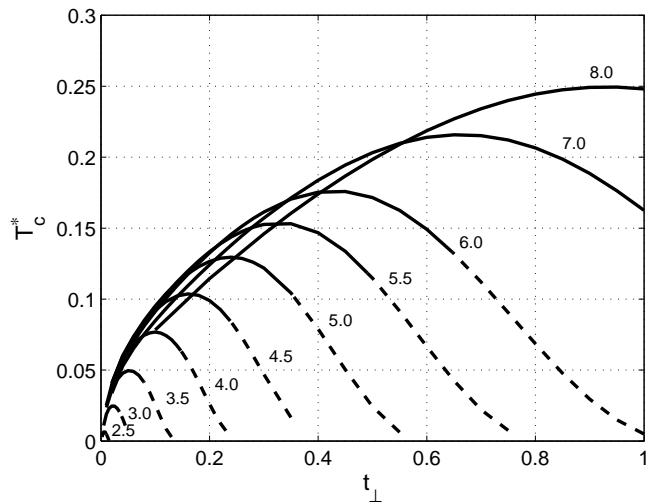}
\caption{Close-packed critical temperature $T^{\ast}_c$ for several $V$ as a function of interplane hopping $t_{\perp}$. $V_{\perp} = 0$, $U = 10$, $A = 0.1$. Numbers by the plots indicate the value of $V$. All quantities are measured in units of $t$. Dashed lines mark the regions where $T^{\ast}_c > \Delta$. }
\label{UV3d:fig:three}
\end{center}
\end{figure}

\section{\label{UV3d:sec:four}
Results
}

\subsection{\label{UV3d:sec:fourone}
Close-packed critical temperature 
}

Using the methods described in section~\ref{UV3d:sec:three}, we solve equation (\ref{3UV2d:eq:four}) for $T^{\ast}_c$ at different model parameters in (\ref{3UV2d:eq:five}). The results for $U = 10$ are shown in figure~\ref{UV3d:fig:three}. As a function of interlayer hopping $t_{\perp}$, $T^{\ast}_c$ shows a pronounced maximum, consistent with the arguments presented in the introduction. At large $t_{\perp}$, kinetic energy is large, pairs are barely formed, their effective radius is large and close-packed density is small. The optimal $T_c$ drops to zero as a result. In the opposite limit of very small $t_{\perp}$, the pairs are stable but increasingly confined within planes. The interplane effective mass goes up, the pairs lose 3D coherence, and $T^{\ast}_c$ drops to zero, see (\ref{3UV2d:eq:three}). Optimal $T^{\ast}_c$ occurs at intermediate interlayer hopping. 

Similarly, there is an optimal attraction $V$. Consider, for example, $t_{\perp} = 0.4$. At $V = 4.5$, the pairs are barely bound, their volume is large, the packing density is small and $T^{\ast}_c$ is small. Increasing $V$ compacts the pairs and increases the packing density. The critical temperature grows until about $V = 7$, after which the competing process of pairs becoming too heavy takes over and $T^{\ast}_c$ falls again. These optimal values of $V$ exceed the threshold of fermion clusterization in the pure 2D $UV$ model~\cite{Kornilovitch2014}. As a result, phase separation is likely to happen at lower $V$ than the maximal $T^{\ast}_c$. For that reason, a search for the absolute maximum of $T^{\ast}_c$ as a function of $V$ and $t_{\perp}$ is not attempted here.         

\begin{figure}[t]
\begin{center}
\includegraphics[width=0.60\textwidth]{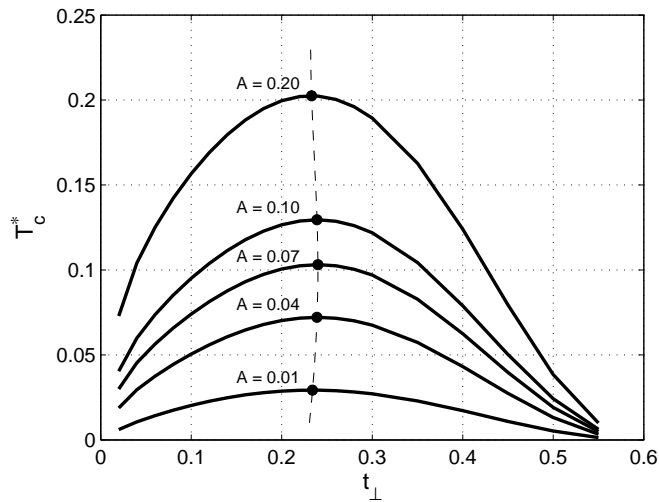}
\caption{Close-packed critical temperature $T^{\ast}_c$ for various coefficients $A$ in (\ref{3UV2d:eq:four}). Circles mark the maxima of curves and the dashed line is an interpolation between them. $U = 10$, $V = 5$, $V_{\perp} = 0$.}
\label{UV3d:fig:four}
\end{center}
\end{figure}

Sensitivity to the close packing parameter $A$ is now discussed. As stated above, the uncertainty in $A$ reflects the lack of an exact criterion for the maximal critical temperature in the absence of a many-body solution. One can only argue that $A$ is probably less than one, while its precise value is unknown. An obvious way of dealing with the uncertainty is to compute $T^{\ast}_c$ for different $A$ and examine variation of optimal parameters. A typical set of curves for $U = 10$ and $V = 5$ is shown in figure~\ref{UV3d:fig:four}. In this example, $A$ is changed between 0.2 and 0.01. The corresponding optimal interlayer hopping is confined between 0.23 and 0.24, i.e. $t_{\perp}$ is robust against variations of $A$. In contrast, the peak critical temperature of each curve varies considerably with $A$. In fact, numerical dependence, $T^{\ast}_{c, max} \propto A^{0.646}$, is exactly what is expected from an ideal Bose gas, with the exponent being close to $2/3$. From the $T_c$ standpoint, variation of $A$ is just a renormalization of the overall boson density. Within the present approach one can only claim the existence of an optimal interlayer hopping and its approximate value: $t_{\perp} \approx 0.23$ for the example in figure~\ref{UV3d:fig:four}. However, one cannot claim an absolute value of $T^{\ast}_{c, max}$ itself.

Self-consistency of preformed pair mechanism requires $T^{\ast}_c$ to be smaller than the pair binding energy $\Delta$. When $T^{\ast}_c > \Delta$, superconductivity crosses over to the BCS regime. These regions are marked in figure~\ref{UV3d:fig:three} by dashed lines. Thus, for the chosen value of $A = 0.1$, optimal parameters ($T^{\ast}_c$ peaks) satisfy $T^{\ast}_c < \Delta$, and the mechanism is self-consistent. One should add that the cross-over points also scale with $A$: smaller $A$ reduce $T^{\ast}_c$ and expand the domain of $T^{\ast}_c < \Delta$. Because of the uncertainty in $A$, the presented argument is only qualitative in nature.   

\begin{figure}[t]
\begin{center}
\includegraphics[width=0.60\textwidth]{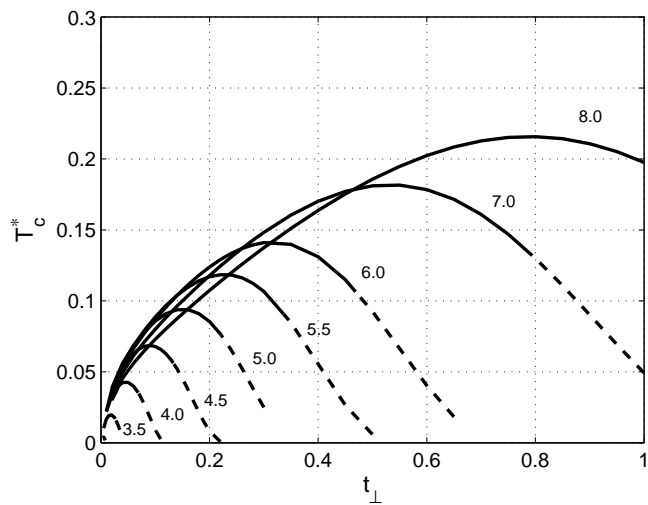}
\caption{Same as figure~\ref{UV3d:fig:three} but for $U = 50$.}
\label{UV3d:fig:five}
\end{center}
\end{figure}

Figure~\ref{UV3d:fig:five} shows $T^{\ast}_c$ vs. $t_{\perp}$ dependence for a large Hubbard repulsion $U = 50$. Compared with $U = 10$, the curves uniformly shift down. This reflects the unbinding action of $U$. For the same $V$ and $t_{\perp}$, a larger $U$ reduces the binding energy and as a result the close packed density of pairs. However, qualitatively curve shapes remain unchanged. The $T^{\ast}_c$ peaks shift to smaller $t_{\perp}$, which reflects the need to reduce the interlayer kinetic energy to compensate a stronger on-site repulsion.

\subsection{\label{UV3d:sec:fourtwo}
Phase diagram 
}

\begin{figure}[t]
\begin{center}
\includegraphics[width=0.60\textwidth]{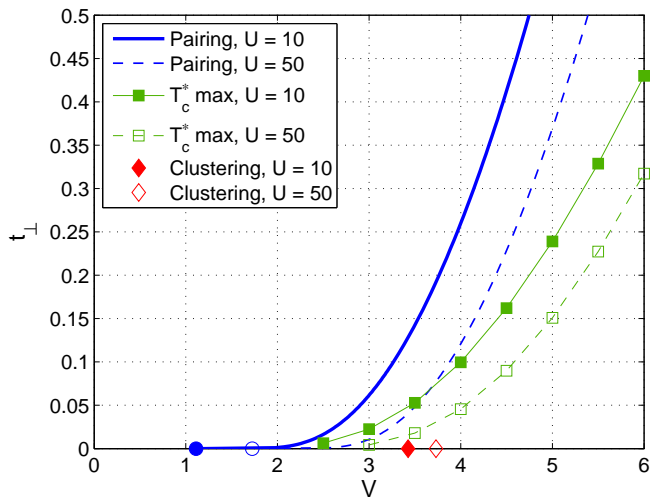}
\caption{(Color online) Phase diagram in $(V,t_{\perp})$ coordinates for $U = 10$ (solid lines, filled symbols) and $U = 50$ (dashed lines, open symbols). $V_{\perp} = 0$. The diamonds on the $V$-axis mark three-fermion clustering thresholds in the pure 2D $UV$ model~\cite{Kornilovitch2014}. }
\label{UV3d:fig:six}
\end{center}
\end{figure}

The results of the preceding sections are summarized in a $(V, t_{\perp})$ phase diagram shown in figure~\ref{UV3d:fig:six}. A pairing line separates the regions of bound and unbound pairs. Above the pairing line, attraction is not strong enough against kinetic energy, pairs do not form and preformed pair superconductivity does not exist. The pairing lines have been computed by solving the eigenvalue equation (\ref{3UV2d:eq:thirteen}) for zero pair momentum ${\bf K}$, to find the value of $V$ for which the binding energy is equal to the noninteracting value, $E = - 8 t - 4 t_{\perp}$. Circles mark the termination points of the lines, determined by the binding condition for a pure 2D $UV$ model (\ref{3UV2d:eq:seven}). Notice how singular the pairing boundary is near $t_{\perp} = 0$. This is because the matrix elements $M_{nml}$ (\ref{3UV2d:eq:fourteen}) are dominated by a logarithmic singularity (which is characteristic of the pure 2D case) only at extremely small $t_{\perp} < 0.001$. At higher $t_{\perp}$, regular contributions are comparable with the logarithmic contributions, which quickly pulls away the solution of (\ref{3UV2d:eq:thirteen}) from its $t_{\perp} = 0$ limit. In other words, pair motion is more 3D like rather than 2D like for all $t_{\perp}$, unless the latter is unphysically small.     

Under the pairing line is the region of preformed pair superconductivity. As shown above, for each $V$ there exists an optimal interlayer hopping at which the close packed critical temperature is maximal. These $t^{\ast}_{\perp}$ are shown as squares. (The connecting lines are guides to the eye only.) In general, $t^{\ast}_{\perp}$ is an increasing function of $V$.   

As can be seen in figures \ref{UV3d:fig:three} and \ref{UV3d:fig:five}, the peak $T^{\ast}_c$ increases with $V$ at least until $V \approx 8$, after which $T^{\ast}_c$ is expected to fall due to a high pair mass. It suggests increasing $V$ as much as possible as a way to boost the critical temperature. This path runs into a difficulty associated with phase separation: any finite range attractive interaction will form multi-fermion clusters in the strong attraction limit. The phase separation threshold in the full tetragonal $UV$ model is unknown at present. One can gain some insight from the recently completed analysis of the pure 2D $UV$ model that corresponds to $t_z = 0$~\cite{Kornilovitch2014}. Three fermion clustering takes place at $V = 3.425$ for $U = 10$ and at $V = 3.730$ at $U = 50$. It is not obvious {\em a priori} how the phase separation boundary behaves as a function of $t_{\perp}$: will it have a logarithmic-like singularity similar to the pairing line, or will it behave more regularly? However, on physical grounds one can expect that finite $t_{\perp}$ will require $V > 3.4$ to form clusters. This leaves preformed pair superconductivity a `region to operate' of at least $2.0 < V < 3.4$ or wider, depending on the intersection location. Within this region, larger $V$ implies larger peak $T^{\ast}_c$. The peak critical temperature can be systematically increased by increasing $V$ and adjusting interlayer hopping to an optimal $t^{\ast}_{\perp}$, until the system runs into phase separation. One arrives at an important conclusion: {\em in the preformed pair mechanism, systems with the highest critical temperatures are always close to phase separation}.  In the Authors' opinion, this is the fundamental reason why so many high-$T_c$ superconductors exhibit a tendency to charge order instabilities including stripes, charge density waves and nematic order~\cite{Tranquada1995,Ghiringhelli2012,Chang2012,Chu2010}.

\subsection{\label{UV3d:sec:five}
Retardation effects
}

In practice, models of the $UV$ form involving an effective instantaneous density-density interaction have their origins in interactions mediated via bosons. All of these interactions are retarded in the sense that when there is interaction between two fermions, the absorption (and therefore scattering) of the mediating boson by a second fermion takes place at a later time than the emission from the first fermion. The aim of this section is to establish that the peak seen in plots of $T_{c}^{\ast}$ vs $t_{\perp}$ is also present when interactions include more realistic retardation effects.

In this section, we use continuous time path-integral Quantum Monte Carlo (QMC) simulations to consider the effects of a retarded interaction described by a Hamiltonian of the form 
\begin{eqnarray} 
H & = & - t \sum_{{\bf m}, {\bf b}, \sigma } 
c^{\dagger}_{{\bf m} \sigma} c_{{\bf m} + {\bf b}, \sigma}  
- t_{\perp} \sum_{{\bf m}, {\bf b}_{\perp}, \sigma } 
c^{\dagger}_{{\bf m} \sigma} c_{{\bf m} + {\bf b}_{\perp}, \sigma}  
\nonumber \\
& & + \sum_{{\bf m} {\bf m}' \sigma } \gamma_{{\bf m}{\bf m}'} 
c^{\dagger}_{{\bf m} \sigma} c_{{\bf m} \sigma} 
( d^{\dagger}_{{\bf m}'} + d_{{\bf m}' } ) 
\nonumber \\
& & + \sum_{{\bf k}} \hbar \omega_{\bf k} \, d^{\dagger}_{{\bf k}} d_{{\bf k}}
    + U \sum_{\bf m} n_{\bf m} ( n_{\bf m} - 1 )  \: ,
\label{3UV2d:eq:twentytwo} 
\end{eqnarray} 
where $d^{\dagger}_{{\bf m} s}$ creates a boson with spin $0$ (e.g. a phonon). For reasons of computational simplicity, only interactions with dispersionless Einstein phonons and infinite repulsive Hubbard $U$ are considered. The level of retardation is refined by the frequency of the boson mode, $\omega_{\bf k} = \omega$ which is taken to be momentum independent (equivalent to site local). For infinite $\omega$, the interaction is instantaneous, with retardation effects increasing as $\omega$ decreases. In the following, $\hbar \omega = 4 t$ is selected to introduce moderate retardation effects. $\gamma_{{\bf m}{\bf m}'}$ represents a long-range interaction. The retarded potential is chosen to have a range of one lattice site, with the form $\gamma_{0, \pm{\bf x}} = \gamma_{0, \pm{\bf y}} = \gamma_{00}/4$ (of the type used in \cite{Davenport2012}, but with the interaction turned off in the $z$-direction). With an appropriate canonical transformation as $\omega \rightarrow \infty$, such a model can be transformed into the $UV$ form, where $U = \infty$ and $V = {\cal V}_{\rm eb} = \sum_{\bf m}\gamma_{0{\bf m}}\gamma_{{\bf m}{\bf x}}/\hbar \omega$ (see e.g. \cite{Hague2007b}). To avoid confusion with the unretarded $UV$ model, a new parameter, ${\cal V}_{\rm ep}$, is introduced. While the simulations here consider phonons, the main difference between interactions with spin-1 bosons that cause spin flips (e.g. magnons) and those that do not (e.g. phonons) is that scattering involving spin-1 bosons is forbidden if both particles occupy the same site. Since the infinite repulsion considered here stops double occupancy of sites, the properties of bound pairs formed from both types of interaction are expected to be broadly similar.

We have already published extensive details of the CTQMC algorithm, so we only discuss differences in measurements here and otherwise refer the reader to \cite{Davenport2012,Kornilovitch1999,Hague2007b,hague1d,hague2d}. The close packed transition temperature is calculated from QMC data using the expression 
\begin{equation}
k_B T^{\ast}_{c} = t \left[ \frac{m^{-1}_{x} m^{-1}_{y} m^{-1}_{z} \left( \frac{t_{\perp}}{t} \right)}
{( \langle \Delta x^2 \rangle + 1 )( \langle \Delta y^2 \rangle + 1 )
 ( \langle \Delta z^2 \rangle + 1 ) } 
\right]^{1/3} .
\label{3UV2d:eq:twentythree}
\end{equation}
Here $m_{i}^{-1}$ are components of the inverse effective mass. $m_{x}^{-1}$ and $m_{y}^{-1}$ are in units of the $xy$ band mass $m_0$ and $m_{z}^{-1}$ is in units of the $z$ band mass $m_{0\perp}$. The ratios $m_{0}/m_{x}$, $m_{0}/m_{y}$ and $m_{0\perp}/m_{z}$ can be computed by QMC as explained elsewhere~\cite{Hague2007b}. The expectation value $\langle \Delta {\bf r}_{i}^2 \rangle$ is calculated by stochastic averaging over an ensemble of imaginary-time fermion paths using an estimator $\Delta {\bf r}_{i}^2 = \beta^{-1} \int_{0}^{\beta}[ \Delta {\bf r}(\tau) \cdot {\bf e}_{i} ]^2 d \tau$, where $\Delta {\bf r}(\tau)$ is the distance between paths at imaginary time $\tau$, and ${\bf e}_{i}$ is a unit vector along direction $i$.

\begin{figure}[t]
\begin{center}
\includegraphics[width=0.60\textwidth]{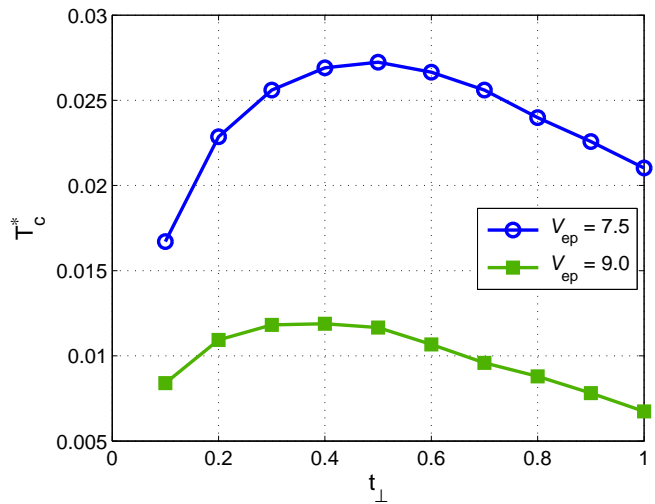}
\caption{(Color online) Transition temperatures, $T^{\ast}_c$, calculated from the effective masses and pair sizes obtained from continuous time QMC and substituted into equation (\ref{3UV2d:eq:twentythree}). The phonon energy is $\hbar \omega = 4 \, t$ and the transition temperature is shown in units of $t$. A peak is clearly visible at intermediate hopping. For the simulation parameters used here, the pair (bipolaron) masses and radii leading to the transition temperatures shown in the figure are already stable at  simulation temperatures of $\beta^{-1} = 0.10 \, t$, since the pair binding energy is significantly greater than $\beta^{-1}$. As such, the figure represents the transition temperature for preformed pair superconductivity. The mass and radius have also been checked for stability at temperatures of $0.05 \, t$, $0.02 \, t$ and $0.01 \, t$ with no significant changes.}
\label{UV3d:fig:seven}
\end{center}
\end{figure}

In the continuous time QMC results shown in figure~\ref{UV3d:fig:seven} for ${\cal V}_{\rm ep} = 7.5$ and ${\cal V}_{\rm ep} = 9.0$, a peak in $T^{\ast}_{c}$ is also visible around $t_{\perp} = 0.5$, similar to the instantaneous $UV$ case. There are some important differences between retarded and instantaneous interactions. Firstly, a retarded interaction leads to self-interaction such that the mass and therefore effective hopping of a {\em single} particle vary with the coupling, whereas these are fixed for the instantaneous coupling. (One should mention that single particle masses in the $z$ direction scale differently to those in the $xy$ plane~\cite{Kornilovitch1999}). This can lead to the peak in $T^{\ast}_c$ being in subtly different positions, but (more importantly) reducing $T^{\ast}_{c}$ at large ${\cal V}_{\rm ep}$ at a faster rate than for the unretarded $UV$ model. Secondly, the QMC calculations are done at finite temperature $\beta^{-1}$, which is then reduced to check that the pair binding radius and mass are unchanged, which is reasonable since at the values of ${\cal V}_{\rm ep}$ considered the binding energy of the pair is significantly larger than the temperature. The finite temperature means that $T^{\ast}_{c}$ estimates are difficult at weak coupling where the pair is only weakly bound. In spite of these differences, it is clear that the interplane hopping, $t_{\perp}$ has a non-trivial effect on the condensation transition temperatures, with a peak clearly visible at intermediate interplane hopping.

\section{\label{UV3d:sec:six}
Summary and conclusions
}

Motivated by the unusually high transition temperatures in layered FeSe and cuprate superconductors, we have studied in this paper the effects of interlayer hopping on the superconducting transition temperatures of systems with local fermion pairs. In particular, our aim was to understand why some of the best superconductors are quasi 2D in spite of an absence of superconducitivity in pure 2D systems. An appropriate theoretical system for studying the effects of interlayer hopping is the tetragonal $UV$ model defined in equation (\ref{3UV2d:eq:five}). There, $V$ and $V_{\perp}$ are attractive pseudopotentials that overcome on-site Hubbard repulsion $U$ and kinetic terms $t$ and $t_{\perp}$ to bind fermions into $s$-wave pairs. On physical grounds, we considered a version of the model with $V_{\perp} = 0$. We have exactly solved the problem of two-fermion pairing using a number of analytical and numerical techniques described in section~\ref{UV3d:sec:three} and the appendices. The pairs evolve from weakly coupled ones, with a small effective mass and large radius at small $V$, to tightly bound ones with a large binding energy, effective mass that scales $\propto V$, and an effective radius of the order of one lattice cell at large $V$. We calculated masses and pair volumes using exact techniques, and used them to determine the BEC temperature of the local pairs.

Central to the present work is the idea that the condensation temperature is maximal when pairs are close packed. The density of close packed pairs is determined by exactly calculating the pair volume. In this way, a maximal density of pairs can be found such that pairs do not overlap. Indeed, so long as the pairs do not overlap, $T_c$ can be increased further by adding more pairs since $T_c \propto \nu^{2/3}_p$. In the opposite limit of pairs overlapping heavily, the additional repulsion brought by the exclusion principle evaporates the pairs and destroys local {\em pair} superconductivity. (Such a system may still exhibit a weakly coupled superconductivity of BCS type.) This reasoning leads to the self-consistency condition (\ref{3UV2d:eq:four}) that defines the close-packed condensation temperature $T^{\ast}_c$ for any given set of model parameters. 

On the basis of (\ref{3UV2d:eq:four}) and (\ref{3UV2d:eq:five}), the effects of interlayer hopping $t_{\perp}$ have been established. Large $t_{\perp} \approx t$ increases the overall kinetic energy of constituent fermions so the pairs cannot bind as easily. In the limit of zero binding energy the effective pair volume diverges, the close packed density goes to zero and $T^{\ast}_c \rightarrow 0$ as a result. At small $t_{\perp} \rightarrow 0$, the pairs are well formed and stable, but lose three dimensional coherence. The $z$-direction effective mass diverges and $T^{\ast}_c \rightarrow 0$ again, in accordance with (\ref{3UV2d:eq:three}). As a result, $T^{\ast}_c$ has a maximum as a function of $t_{\perp}$. Specific examples of $T^{\ast}_c (t_{\perp})$ dependencies are shown in figures \ref{UV3d:fig:three} and \ref{UV3d:fig:five} for the instantaneous $UV$ potential and in figure~\ref{UV3d:fig:seven} for a retarded attractive potential. {\em The existence of local preformed pairs naturally predicts the existence of optimal interlayer hopping, and this optimal value is less than one, so maximal condensation temperatures are expected in strongly anisotropic quasi two dimensional systems.}   This is a strong indicator that anisotropy is an important feature in high temperature supercondutors.

Another important conclusion can be reached by analyzing the effect of attraction $V$ on the close packed critical temperature. Again, $T^{\ast}_c(V)$ exhibits a broad maximum: at small $V$ the pairs do not bind and the close packed density is zero, whereas at large $V$ the pairs are well bound but their masses are large, causing $T^{\ast}_c$ to drop in accordance with (\ref{3UV2d:eq:three}). However, the optimal $V$ falls in the region of phase separation. For superconductivity to be stable, $V$ must be less than optimal. On this side of the curve, optimal $T^{\ast}_c$ can be systematically raised by increasing $V$ until the system hits phase separation. At large $V$ where phase separation occurs, pairs become bound states of three or more particles (which in dense systems will become the precursors for states such as stripes and nematic order). The mass of such bound states is significantly greater than the pair mass. Any states with an odd number of particles are fermions and do not directly condense. Any phase separated states with an even number of particles could in principle Bose condense, however the mass of such particles would be many orders of magnitude higher than that of the pair particles and the BEC temperature would be so low that superconductivity is essentially destroyed. As the clusters become macroscopic states (e.g. stripes and nematic order) then the preformed pair superconductivity described in this paper would be completely destroyed. Within the preformed pair mechanism, the highest pair condensation temperatures are always close to phase separation and as such materials with this mechanism are expected to exhibit a variety of charge order instabilities including stripes, charge density waves and nematic order~\cite{Tranquada1995,Ghiringhelli2012,Chang2012,Chu2010}. This may be the reason why high temperature superconductivity often occurs in the neighbourhood of phase separation phenomena. Figure~\ref{UV3d:fig:six} shows the summary phase diagram of the tetragonal $UV$ model.  

Further work will involve analysis of clustering and phase separation. Rigorous analysis of phase separation in $UV$ lattice models is difficult, but some progress was recently made~\cite{Kornilovitch2014}. At strong enough $V$ the particles form three fermion clusters, four fermion clusters and so on. The system phase separates and becomes a poor metal, destroying superconductivity. As a result, optimal preformed pair superconductivity is never far from phase separation. This delicate balance presents a major challenge for any analytical treatment, and it is expected that advanced numerical techniques are needed to determine the limits of local pair superconductivity.

\ack

The authors wish to thank V Bulatov and A Davenport for useful discussions.

\appendix

\section{ \label{UV3d:sec:appa}
Evaluation of matrix elements $M_{nml}$ 
}

Matrix elements $M_{nml}$ in Eq.~(\ref{3UV2d:eq:fourteen}) are given by generalized Watson integrals 
\begin{equation}
M_{nml} = \int^{\pi}_0 \!\!\!\! \int^{\pi}_0 \!\!\!\! \int^{\pi}_0  
\frac{dx dy dz}{\pi^3} \frac{\cos{nx} \cos{my} \cos{lz}}
{\vert E \vert - a \cos{x} - b \cos{y} - c \cos{z}}  ,
\label{3UV2d:eq:appone}
\end{equation}
where $a \equiv 4 t \cos{(K_x/2)}$, $b \equiv 4 t \cos{(K_y/2)}$, and $c \equiv 4 t_{\perp} \cos{(K_z/2)}$. Note that although the original tetragonal $UV$ model is isotropic in the $xy$ plane, nonzero pair momentum ${\bf K}$ breaks that symmetry, thereby requiring orthorhombic $M_{nml}$. The integrals (\ref{3UV2d:eq:appone}), also known as lattice Green functions, have been researched for more than 70 years~\cite{Zucker2011}. Despite impressive progress with analytical integration achieved in the last decade~\cite{Joyce2002,Joyce2003,Joyce2006,Guttmann2010} no closed-form expression for orthorhombic $M_{nml}$ with $a \neq b \neq c$ exists. However, known analytical results for symmetric corner cases can be used to gauge the accuracy of numerical procedures.        

Purely numerical evaluation of $M_{nml}$ is impractical either. Near the pair formation threshold, $\vert E \vert \rightarrow (a + b + c)$, integrands become singular but integrals remain finite. Proper handling of the singularity requires nonuniform 3D meshes that are error prone. Additionally, for strong hopping anisotropies, $t_{\perp} \ll t$, the integrals diverge logarithmically. To capture the latter, the mesh must be $\vert E \vert$ and $t_{\perp}$ dependent, further complicating the matter.    

A practical approach consists of carrying two integrations analytically and leaving the third one to numerics. Handling singularities in 1D integrals is much easier and often included in standard numerical packages. The first integration in (\ref{3UV2d:eq:appone}) is elementary but the second is not. The second integration is in fact a transformation to complete elliptic integrals of the first and second kind ${\rm K}(\kappa)$ and ${\rm E}(\kappa)$. For the purposes of this paper it is sufficient to know two auxiliary integrals~\cite{Prudnikov1986}   
\begin{eqnarray}
{\cal M}_{00} & = & \int^{\pi}_0 \!\!\!\! \int^{\pi}_0  
\frac{dx dy}{\pi^2} \frac{1}{ {\cal E} - a \cos{x} - b \cos{y} } 
\nonumber \\
& = & \int^{\infty}_0 du \, e^{- {\cal E} u} I_0(au) I_0(bu) = 
\frac{\kappa}{\pi \sqrt{ab}} \, {\rm K}(\kappa) \: ,
\label{3UV2d:eq:apptwo}
\end{eqnarray}
\begin{eqnarray}
{\cal M}_{11} & = & \int^{\pi}_0 \!\!\!\! \int^{\pi}_0  
\frac{dx dy}{\pi^2} \frac{\cos{x} \cos{y}}{ {\cal E} - a \cos{x} - b \cos{y} } = 
\int^{\infty}_0 du \, e^{- {\cal E} u} I_1(au) I_1(bu) 
\nonumber \\
& = & \frac{1}{\pi \kappa \sqrt{ab}} 
\left[ ( 2 - \kappa^2 ) {\rm K}(\kappa) - {\rm E}(\kappa) \right] \: ,
\label{3UV2d:eq:appthree}
\end{eqnarray}
where $I_{0,1}(u)$ are modified Bessel functions of order 0 and 1, ${\cal E} > a + b$, and
\begin{equation}
\kappa = \sqrt{\frac{4 a b}{ {\cal E}^2 - (a - b)^2 }} \: .
\label{3UV2d:eq:appfour}
\end{equation}
Applying (\ref{3UV2d:eq:apptwo}) and (\ref{3UV2d:eq:appthree}) to (\ref{3UV2d:eq:appone}), one can derive working expressions for all $M_{nml}$. For example,  
\begin{equation}
M_{020} =  \frac{1}{\pi^2 \sqrt{ac}} \int^{\pi}_0 dy \, \cos{(2y)} \, \kappa_y(y) 
{\rm K} \left[ \kappa_y(y) \right] \: ,
\label{3UV2d:eq:appfive}
\end{equation}
\begin{equation}
\kappa_y(y) = \sqrt{\frac{4 a c}{ \left( \vert E \vert - b \cos{y} \right)^2 - (a - c)^2 }} \: .
\label{3UV2d:eq:appsix}
\end{equation}
The remaining 1D numerical integration is fast enough to enable effective eigenvalue search for (\ref{3UV2d:eq:thirteen}). In two special cases: (i) $a = b = c$, $n,m,l = 0,1,2$ and arbitrary $\vert E \vert$, and (ii) $a = b \neq c$, $n = m = l = 0$ and arbitrary $\vert E \vert$, numerical integration was validated against the analytical results by Joyce et al~\cite{Joyce2002,Joyce2003}. Close to logarithmic divergence, $t_{\perp} \rightarrow 0$ and $\vert E \vert \rightarrow (a + b + c)$, the procedure was stabilized further by computing differences $M_{nml} - M_{000}$ that involved only nonsingular functions. The base integral $M_{000}$ was computed three times by applying three different versions of (\ref{3UV2d:eq:apptwo}). A match between the three values within an integration tolerance of order $10^{-12}$ served as an internal consistency check for the entire method.

\begin{figure}[t]
\begin{center}
\includegraphics[width=0.60\textwidth]{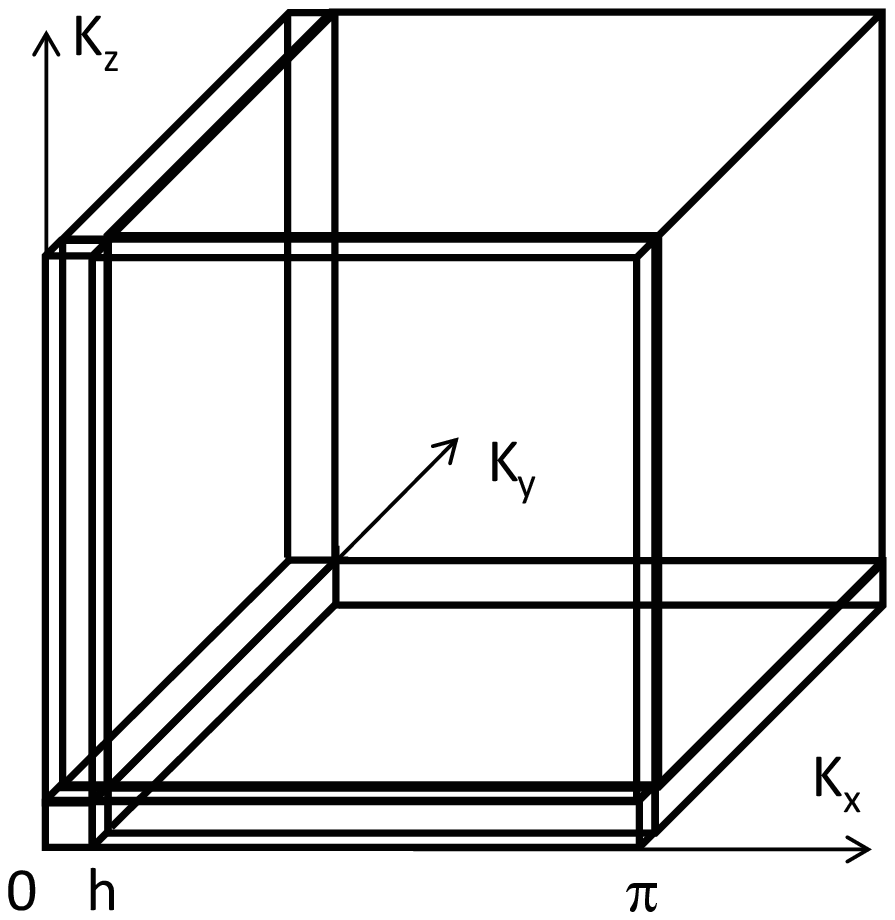}
\caption{Subdomains for Brillouin zone integration in (\ref{3UV2d:eq:four}).}  
\label{UV3d:fig:eight}
\end{center}
\end{figure}

\section{ \label{UV3d:sec:appb}
Subdomain integration in the Bose integral (\ref{3UV2d:eq:four}) 
}

The irreducible integration domain $0 \leq K_{i} \leq \pi$ is split into eight subdomains as illustrated in figure~\ref{UV3d:fig:eight}. The parameter $h$ is of order 0.01. The eight subdomains are:

(i) One cube $h \leq K_{i} \leq \pi$, to which uniform meshing and 3D Simpson integration is applied.

(ii) Three plane-like square prisms running along coordinate planes. For example, for the $(xy)$ prism, $h \leq K_{x,y} \leq \pi$ and $0 \leq K_{z} \leq h$. Let 
\begin{equation}
\phi({\bf K}) = \exp{ \left\{ \frac{ E({\bf K}) - E(0) }{ k_B T^{\ast}_c} \right\} } - 1 \: .
\label{3UV2d:eq:fifteen}
\end{equation}
Variation of integrand $\phi^{-1}({\bf K})$ with $K_{z}$ is replaced with a Taylor expansion, integration over $K_z$ is performed analytically and the result is expressed via integrand values (and, eventually, pair energy $E$) at the top and bottom faces of the prism:
\begin{eqnarray}
I_{xy} = && \frac{2h}{3} \int^{\pi}_h \!\!\! \int^{\pi}_h dK_x dK_y \, \frac{1}{\phi( K_x, K_y, 0 )} + 
\nonumber \\
         && \frac{h}{3}  \int^{\pi}_h \!\!\! \int^{\pi}_h dK_x dK_y \, \frac{1}{\phi( K_x, K_y, h )} \: .
\label{3UV2d:eq:sixteen}
\end{eqnarray}
A 2D Simpson rule is applied to the remaining integrals, which are nonsingular. 

(iii) Three rod-like square prisms running along coordinate axes. For example, for the $(y)$ prism, $h \leq K_{y} \leq \pi$ and $0 \leq K_{x,z} \leq h$. Variation of the integrand with $K_{x,z}$ is replaced with a Taylor expansion, integration is done analytically and the result is expressed via $\phi$ on prism edges:  
\begin{equation}
I_{y} = \frac{h^2}{3} \!\! \int^{\pi}_h \!\! dK_y \!\! \left[ 
\frac{1}{\phi( 0, K_y, 0 )} \! + \! \frac{1}{\phi( h, K_y, 0 )} \! + \! 
\frac{1}{\phi( 0, K_y, h )} \right] .
\label{3UV2d:eq:seventeen}
\end{equation}
A 1D Simpson rule is then applied to the remaining $K_y$ integral. Similar expressions are developed for the two other rod integrals $I_{x}$ and $I_{z}$. 

(iv) One cube $0 \leq K_{x,y,z} \leq h$ which remains singular. Making use of $h \ll 1$, $\phi({\bf K})$ is replaced with a parabolic approximation, and integration over $K_z$ is done analytically. The remaining double integral 
\begin{equation}
I_{0} = \frac{h^2}{\sqrt{\phi(0,0,h)}} \int^{h}_{0} \!\!\! \int^{h}_{0} 
\frac{dK_x dK_y}{w} \, \arctan {\frac{h \sqrt{\phi(0,0,h)}}{w}} \: ,
\label{3UV2d:eq:eighteen}
\end{equation}
where
\begin{equation}
w(K_x,K_y) = \sqrt{\phi(h,0,0) K^2_x + \phi(0,h,0) K^2_y} \: , 
\label{3UV2d:eq:nineteen}
\end{equation}
can be evaluated as repeated one by conventional numerical methods without difficulty. The full right-hand-side of (\ref{3UV2d:eq:four}) is given by the sum of all 8 contributions divided by $\pi^3$.

\section{ \label{UV3d:sec:appc}
Calculation of effective pair radius 
}

The starting point is equation (\ref{3UV2d:eq:eight}). Setting the pair momentum to zero, ${\bf k}_1 = - {\bf k}_2 \equiv {\bf k}$, and Fourier transforming to real space the wave function reads
\begin{equation}
\psi_{{\bf K} = {\bf 0}}( {\bf r}_1, {\bf r}_2 ) = \frac{1}{N} \sum_{\bf b} G_{\bf b} \sum_{\bf k} 
\frac{e^{i {\bf k} ( {\bf r}_1 - {\bf r}_2 + {\bf b} ) }}{E - 2 \varepsilon({\bf k}) } \: .
\label{3UV2d:eq:appbone}
\end{equation}
The nearest-neighbor vectors ${\bf b}$ and quantities $G_{\bf b}$ are listed in the main text after (\ref{3UV2d:eq:twenty}). In the following, the subscript ${\bf K} = {\bf 0}$ will be omitted and the relative vector $({\bf r}_1 - {\bf r}_2)$ will be represented via its lattice coordinates $(n,m,l)$. Substituting here the one-particle spectrum (\ref{3UV2d:eq:six}), replacing the energy denominator with an integral using the identity $x^{-1} = \int^{\infty}_0 d\alpha \exp{(-\alpha x)}$, and making use of the definitions of modified Bessel functions, one obtains 
\begin{equation}
\psi( n,m,l ) = - \sum_{\bf b} G_{\bf b} \int^{\infty}_0 d\alpha \, e^{ - \alpha \vert E \vert } \,
I_{n + b_x}(4 \alpha \, t) \, I_{m + b_y}(4 \alpha \, t) \, I_{l + b_z}(4 \alpha \, t_{\perp}) \: .
\label{3UV2d:eq:appbtwo}
\end{equation}
To compute the mean squared interparticle distance the wave function needs to be normalized. For example, the mean squared $x$-distance is given by  
\begin{equation}
\langle ( x_1 - x_2 )^2 \rangle = \langle n^2 \rangle = 
\frac{ \sum^{\infty}_{n,m,l = -\infty} n^2 \vert \psi( n,m,l ) \vert^2 }
     { \sum^{\infty}_{n,m,l = -\infty}     \vert \psi( n,m,l ) \vert^2 } \equiv
\frac{R_x}{Q}      \: .
\label{3UV2d:eq:appbthree}
\end{equation}
The normalization integral $Q$ can be simplified by making use of the addition identity
\begin{equation}
\sum^{\infty}_{n = -\infty} I_{n + p}(\zeta) I_{n + p'}(\zeta')     = 
\sum^{\infty}_{n = -\infty} I_{n}(\zeta)     I_{p - p' - n}(\zeta') = 
I_{p - p'}( \zeta + \zeta' )      \: .
\label{3UV2d:eq:appbfour}
\end{equation}
Repeated application of (\ref{3UV2d:eq:appbfour}) reduces the normalization integral to a double integral
\begin{eqnarray}
Q & = & \sum_{{\bf b}{\bf b}'} G_{\bf b} G_{{\bf b}'} 
\int^{\infty}_0 \!\!\!\! \int^{\infty}_0 d\alpha \, d\alpha' 
e^{- ( \alpha + \alpha' ) \vert E \vert}     \times 
\nonumber \\
  & \times & I_{b_x - b'_x}[ 4t        (\alpha + \alpha') ] 
\cdot I_{b_y - b'_y}[ 4t(\alpha + \alpha') ] \cdot I_{b_z - b'_z}[ 4t_{\perp}(\alpha + \alpha') ] \: .
\label{3UV2d:eq:appbfive}
\end{eqnarray}
Going over to new variables, $\xi = \alpha + \alpha'$ and $\eta = \alpha - \alpha'$, and integrating over $\eta$, one obtains
\begin{equation}
Q = \sum_{{\bf b}{\bf b}'} G_{\bf b} G_{{\bf b}'} 
\int^{\infty}_0 d\xi \, \xi e^{- \xi \vert E \vert} \cdot I_{b_x - b'_x}( 4t \xi ) 
\cdot I_{b_y - b'_y}( 4t \xi ) \cdot I_{b_z - b'_z}( 4t_{\perp} \xi ) \: .
\label{3UV2d:eq:appbsix}
\end{equation}
This expression defines function $T^{i}_{{\bf b}{\bf b}'}(\xi)$ appearing in the denominator of (\ref{3UV2d:eq:twenty}).  

Shifting now to the numerator of (\ref{3UV2d:eq:appbthree}) and applying the addition theorem to two pairs of $I$, $R_x$ assumes the form
\begin{eqnarray}
R_x & = & \sum_{{\bf b}{\bf b}'} G_{\bf b} G_{{\bf b}'} 
\int^{\infty}_0 \!\!\!\! \int^{\infty}_0 d\alpha \, d\alpha' 
e^{- ( \alpha + \alpha' ) \vert E \vert} \times
\nonumber \\
 & \times & I_{b_y - b'_y}[ 4t(\alpha + \alpha') ] 
\cdot I_{b_z - b'_z}[ 4t_{\perp}(\alpha + \alpha') ] \cdot X_{{\bf b}{\bf b}'}(\alpha,\alpha') \: ,
\label{3UV2d:eq:appbseven}
\end{eqnarray}
\begin{equation}
X_{{\bf b}{\bf b}'}(\alpha,\alpha') = \sum^{\infty}_{n = - \infty} n^2 
\cdot I_{n + b_x}( 4t \alpha ) I_{n + b'_x}( 4t \alpha' ) \: . 
\label{3UV2d:eq:appbeight}
\end{equation}
The latter sum can be calculated by: (i) expressing both $I$s as integrals on the interval $[-\pi,\pi]$; (ii) integrating by parts twice to absorb the factor $n^2$; (iii) applying the addition theorem (\ref{3UV2d:eq:appbfour}); (iv) applying recurrence relations for the Bessel functions. Then, after some straightforward algebra the result is
\begin{eqnarray}
\hspace{-1.0cm}
X_{{\bf b}{\bf b}'}(\alpha,\alpha') & = & \frac{ ( \alpha b'_x + \alpha' b_x )^2 }{ ( \alpha + \alpha' )^2 } 
\cdot I_{b_x - b'_x}[ 4t ( \alpha + \alpha' ) ] 
\nonumber \\
 & + & \frac{1}{2} (4t) \frac{ \alpha \alpha' }{ \alpha + \alpha' } 
\left\{ I_{b_x - b'_x + 1}[ 4t ( \alpha + \alpha' ) ] + 
        I_{b_x - b'_x - 1}[ 4t ( \alpha + \alpha' ) ]   \right\} .         
\label{3UV2d:eq:appbnine}
\end{eqnarray}
After substituting (\ref{3UV2d:eq:appbnine}) in (\ref{3UV2d:eq:appbseven}), changing variables to $\xi$ and $\eta$, integrating over $\eta$, and simplifying, $R_x$ finally becomes
\begin{eqnarray}
\hspace{-1.0cm}
R_x = & & \frac{1}{3} \sum_{{\bf b}{\bf b}'} G_{\bf b} G_{{\bf b}'} 
\int^{\infty}_0 d\xi  e^{- \xi \vert E \vert}    
\cdot I_{b_y - b'_y}( 4t \xi ) \cdot I_{b_z - b'_z}( 4t_{\perp} \xi ) 
\\
& & \times \left\{ ( b^2_x + b_x b'_x + b'^2_x ) \, \xi I_{b_x - b'_x}( 4t \xi ) + 
t \xi^2 \left[ I_{b_x - b'_x + 1}( 4t \xi ) + I_{b_x - b'_x - 1}( 4t \xi ) \right] \right\} \: ,
\nonumber
\label{3UV2d:eq:appbten}
\end{eqnarray}
which defined function $S^{x}_{{\bf b}{\bf b}'}(\xi)$ appearing in the numerator of (\ref{3UV2d:eq:twenty}). Expressions for $S^{y}_{{\bf b}{\bf b}'}$ and $S^{z}_{{\bf b}{\bf b}'}$ can be obtained from (\ref{3UV2d:eq:appbten}) by cyclic permutation of $b_x$, $b_y$ and $b_z$ in the indices of the Bessel functions and of $t$, $t$ and $t_{\perp}$ in their arguments.


\section*{References}

\begin{thebibliography}{99}

\bibitem{Dagotto2013}
Dagotto E 2013 {\em Rev. Mod. Phys.} {\bf 85}, 849

\bibitem{He2013}
He S 2013 {\em Nature Materials} {\bf 12}, 605

\bibitem{Ogg1946}
Ogg Jr R A 1946 {\em Phys. Rev.} {\bf 69} 243

\bibitem{Schafroth1954}
Schafroth M R 1954 {\em Phys. Rev.} {\bf 100} 463

\bibitem{Schafroth1957}
Schafroth M R, Butler S T and Blatt J M 1957 {\em Helv. Phys. Acta} {\bf 30} 93

\bibitem{Bogoliubov1970}
Bogoliubov N N 1970 {\em Quasi-Averages in Problems of Statistical Physics} (in {\em Lectures on Quantum Statistics}, vol 2) (Gordon and Breach, New York) 1.

\bibitem{Micnas1990}
Micnas R, Ranninger J and Robaszkiewicz S 1990
{\em Rev. Mod. Phys.} {\bf 62} 113 

\bibitem{Alexandrov1994}
Alexandrov A S and Mott N F  1994 {\em High-Temperature Superconductors and other Superfluids} (Taylor \& Francis, London)

\bibitem{Salje1984}
Salje E and G\"uttler B 1984 {\em Phil. Mag. B} {\bf 50} 607

\bibitem{Kyung1998}
Bumsoo Kyung, Klepfish E G and Kornilovitch P E 1998
{\em Phys. Rev. Lett.} {\bf 80} 3109

\bibitem{Alexandrov1993}
Alexandrov A S and Mott N F 1993
{\em Supercond. Sci. Technol.} {\bf 6} 215

\bibitem{Alexandrov2011}
Alexandrov A S 2011
{\em Phys. Scr.} {\bf 83} 038301

\bibitem{Kornilovitch2014}
Kornilovitch P E 2014
{\em Phys. Rev. Lett.} {\bf 112} 077202

\bibitem{Scalapino2012}
Scalapino D J 2012
{\em Rev. Mod. Phys.} {\bf 84} 1383

\bibitem{Mattis1986}
Mattis D C 1986 {\em Rev. Mod. Phys.} {\bf 58} 361

\bibitem{Kornilovitch1995}
Kornilovitch P E 1995 (in {\em Polarons and Bipolarons in High-$T_c$ Superconductors and Related Materials}
edited by Salje E K H, Alexandrov A S and Liang W Y) (Cambridge University Press) 367
  
\bibitem{Lin1991}
Lin H Q 1991 {\em Phys. Rev. B} {\bf 44} 4674

\bibitem{Petukhov1992}
Petukhov A G, Gal\'an J and Verg\'es J A 1992 {\em Phys. Rev. B} {\bf 46} 6212

\bibitem{Kornilovitch2004}
Kornilovitch P E 2004 {\em Phys. Rev. B} {\bf 69} 235110

\bibitem{Davenport2012}
Davenport A R, Hague J P and Kornilovitch P E 2012
{\em Phys. Rev. B} {\bf 86} 035106 

\bibitem{Zucker2011}
Zucker I J 2011
{\em J. Stat. Phys.} {\bf 145} 591 

\bibitem{Joyce2002}
Joyce G S 2002 {\em J. Phys. A: Math. Gen.} {\bf 35} 9811

\bibitem{Joyce2003}
Joyce G S, Delves R T and Zucker I J 2003 {\em J. Phys. A: Math. Gen.} {\bf 36} 8661 

\bibitem{Joyce2006}
Delves R T and Joyce G S 2006 {\em J. Phys. A: Math. Gen.} {\bf 39} 4119 

\bibitem{Guttmann2010}
Guttmann A J 2010 J. {\em Phys. A: Math. Theor.} {\bf 43} 305205

\bibitem{Capogrosso2007}
Capogrosso-Sansone B, Prokof'ev N V and Svistunov B V 2007 {\em Phys. Rev. B} {\bf 75} 134302  

\bibitem{Yukalov2009}
Yukalov V I 2009 {\em Laser Physics} {\bf 19} 1 

\bibitem{Kleinert2014}
Kleinert H, Narzikulov Z and Rakhimov A 2014
{\em J. Stat. Mech.} P01003 

\bibitem{Tranquada1995}
Tranquada J M, Sternlieb B J, Axe J D, Nakamura Y and Uchida S 1995
{\em Nature} {\bf 375} 561 

\bibitem{Ghiringhelli2012}
Ghiringhelli G, Le Tacon M, Minola M, Blanco-Canosa S, Mazzoli C, Brookes N B, De Luca G M, Frano A, Hawthorn D G, He F, Loew T, Moretti Sala M, Peets D C, Salluzzo M, Schierle E, Sutarto R, Sawatzky G A, Weschke E, Keimer B and Braicovich L 2012 {\em Science} {\bf 337} 821

\bibitem{Chang2012}
Chang J, Blackburn E, Holmes A T, Christensen N B, Larsen J, Mesot J, Ruixing Liang, Bonn D A, Hardy W N, Watenphul A, v Zimmermann M, Forgan E M and Hayden S M 2012 {\em Nature Physics} {\bf 8} 871

\bibitem{Chu2010}
Chu J-H, Analytis J G, De Greve K, McMahon P L, Islam Z, Yamamoto Y and Fisher I R 2010
{\em Science} {\bf 329} 824 

\bibitem{Kornilovitch1999}
Kornilovitch P E 1999 {\em Phys. Rev. B} {\bf 59} 13531

\bibitem{Hague2007b}
Hague J P, Kornilovitch P E, Samson J H and Alexandrov A S 2007 
{\em J. Phys.: Condens. Matter} {\bf 19} 255214

\bibitem{hague1d}
Hague J P and Kornilovitch P E 2009
{\em Phys. Rev. B} {\bf 80} 054301

\bibitem{hague2d}
Hague J P and Kornilovitch P E 2010
{\em Phys. Rev. B} {\bf 82} 094301

\bibitem{Prudnikov1986}
Prudnikov A P, Brychkov Ya A and Marichev O I 1986
{\em Integrals and Series, Special Functions} (Gordon and Breach) vol 2  



\end{thebibliography}
\end{document}